\documentstyle[12pt]{article}

\begin{document}

\author{M.\ Apostol  \\ 
Department of Theoretical Physics,\\Institute of Atomic Physics,
Magurele-Bucharest MG-6,\\POBox MG-35, Romania\\e-mail: apoma@theor1.ifa.ro}
\title{Some remarks on the two-electron atom}
\date{J.\ Theor.\ Phys. {\bf 10} (1995)}
\maketitle

\begin{abstract}
New, approximate, two-electron wavefunctions are introduced for the
two-electron atoms (cations), which account remarkably well for the
ground-state energies and the lowest- excited states (where available). A
new scheme of electronic configurations is also proposed for the
multi-electron atoms.
\end{abstract}

Following Bohr,\cite{Bohr} the electronic structure of the atoms is
currently described by means of the central self-consistent field, based on
the Hartree-Fock equations.\cite{Hartree}$^{,}$\cite{Fock} Although this
mean-field doctrine has long served the interpretation of the atomic spectra,%
\cite{Condon} its central assumption\thinspace \thinspace of single-electron
wavefunctions has never been given an a priori legitimacy. In fact,
correlated ground-state wavefunctions have been worked out in great detail
for the $He$ atom and, generally, for the two-electron atom.\cite{Hylleraas}$%
^{-}$\cite{Pekeris} In spite of the fact that these very elaborate
techniques, which are variational in essence, produce impressively accurate
numerical results for the ground-state energies, they look, however, rather
arbitrary, and can not be extended easily to multi-electron atoms. We
introduce in this paper new, approximate, two-electron wavefunctions for the
two-electron atom, which give the ground-state and the lowest-excited states
energies with an accuracy which may be regarded as being remarkable for the
simplicity of the approach. We propose also a new general scheme of
electronic configurations, which may constitute a reasonable starting point
for understanding the nature of the electronic structure of the
multi-electron atoms.

If not specified otherwise, we use throughout this paper the atomic units $%
a_H=\hbar ^2/me^2=0.53\;\AA $ (Bohr radius) and $e^2/a_H=27.2\;eV$ (twice
the rydberg), which render the square of the electron charge $e^2=1$; in
addition, we set the Planck's constant $\hbar =1$, so that the electron mass 
$m=1$. We start with considering two electrons, denoted by $1$ and $2$,
placed at ${\bf r}_1$ and, respectively, ${\bf r}_2$ in the Coulomb field of
a nuclear charge $Z$. We neglect the center-of-mass corrections, the
spin-orbit coupling, as well as all the other relativistic corrections; they
can be treated perturbationally. Under these assumptions the hamiltonian of
the two-electron atom is given by 
\begin{equation}
\label{one}H=\frac 12p_1^2+\frac 12p_2^2-\frac Z{r_1}-\frac Z{r_2}+\frac
1{\mid {\bf r}_1-{\bf r}_2\mid }\;\;\;, 
\end{equation}
where ${\bf p}_{1,2}$ are the momenta of the two electrons. We shall treat
first the electron affinity of hydrogen, {\it i.e}. the ground-state of the $%
H^{-}$ anion ($Z=1$), and thereafter we shall proceed to the atoms (cations)
with $Z\geq 2$.

If we put ${\bf r}_1=-{\bf r}_2$ in $\left( 1\right) $ (for $Z=1$) we get a
one-particle hamiltonian for a particle of mass $1/2$, moving in a Coulomb
field of charge $3/2$; its ground-state energy is $E_0^{(0)}=-9/16=-15.3\;eV$%
, which amounts to an electron affinity $A^{(0)}=1.7\;eV$.This value is
close to the experimental value $A=0.76\;eV$.\cite{Natl} The zeroth order
approximation $E_0^{(0)}$ must be corrected by higher-order contributions of
the Coulomb repulsion $V=1/\mid {\bf r}_1-{\bf r}_2\mid $ between the two
electrons. For a fixed ${\bf r}_1$ we can see that this Coulomb repulsion
describes, to the first order of approximation, harmonic oscillations of $%
{\bf r}_2=(r_2,\theta _2,\varphi _2)$ along $\theta _2$, around ${\bf r}_2=-%
{\bf r}_1$. The corresponding harmonic-oscillator potentials have, however,
various curvatures, and we should define an average potential. Setting to
describe the small oscillations of ${\bf r}_2$ by the displacements ${\bf %
\rho }$, {\it i.e}. putting ${\bf r}_2=-{\bf r}_1+{\bf \rho }$, we may
expand the Coulomb repulsion $V$ as 
\begin{equation}
\label{two}V=\frac 1{\mid {\bf r}_1-{\bf r}_2\mid }=\frac 1{2r}+\frac \rho
{4r^2}P_1\left( \cos \theta \right) +...\;\;\;, 
\end{equation}
where $r=\mid {\bf r}_1\mid =\mid {\bf r}_2\mid $, $P_1$ is the Legendre
polynomial of the first rank, and $\theta $ is the angle between ${\bf \rho }
$ and ${\bf r}_1$. We define the average potential $\overline{V}$ by $%
\overline{V}=\left( \overline{V^2}\right) ^{1/2}$, where the average in
paranthesis is taken over all the orientations of ${\bf \rho }$, so that, we
get in the first approximation 
\begin{equation}
\label{three}\overline{V}=\frac 1{2r}+\frac{\rho ^2}{48r^3}+...\;\;\;. 
\end{equation}
The first term in $\left( 3\right) $ has already been included in computing
the zeroth order approximation; the second term in $\left( 3\right) $ is a
harmonic-oscillator potential, which is completely determined by replacing $%
r $ by its average value over the zeroth order ground-state $\overline{r}=2$%
. We obtain therefore the harmonic-oscillator frequency $\varpi =\left(
24r^3\right) ^{-1/2}=1.96\;eV$, and the ground-state energy corrected by the
zero-point oscillations $E_0=E_0^{(0)}+\frac 12\varpi =-14.32\;eV$; this
amounts to an electron affinity $A=0.72\;eV$ which is in excellent agreement
with the experimental value $0.76\;eV$. We remark that the oscillations of
the second electron are along $\theta _2$ , so that they correspond to only
one degree of freedom.

For $Z\geq 2$ we define $r=\min \left( r_1,r_2\right) $ and $R=\max \left(
r_1,r_2\right) $ and separate the hamiltonian $\left( 1\right) $ as follows: 
\begin{equation}
\label{four}H=H_Z+H_{Z_1}+V\;\;\;, 
\end{equation}
\begin{equation}
\label{five}H_Z=\frac 12p^2-\frac Zr\;\;\;, 
\end{equation}
\begin{equation}
\label{six}H_{Z_1}=\frac 12P^2-\frac{Z_1}R\;\;\;, 
\end{equation}
\begin{equation}
\label{seven}V=\frac 1{\mid {\bf r}_1-{\bf r}_2\mid }-\frac 1R\;\;\;, 
\end{equation}
where $Z_1=Z-1$ and ${\bf p},\;{\bf P}$ are the momenta of the electrons
placed at ${\bf r}$ and, respectively, ${\bf R}$. This separation
corresponds to the classical concept of screening. The two hamiltonians $H_Z$
and $H_{Z_1}$ have a hydrogen-like energy spectrum; the ground-state
wavefunction is given by $\Psi _0({\bf r}_1,{\bf r}_2)=C_0\Psi _{100}^{(Z)}(%
{\bf r)}\Psi _{100}^{(Z_1)}({\bf R)}$, where $\Psi _{100}^{(Z,Z_1)}$ are the
corresponding ground-state wavefunctions of the hydrogen-like atom
(normalized to unity) and $C_0$ is a normalization constant.This
wavefunction corresponds to an electron configuration which may be denoted
by $\left( 100\right) \times (100)$, and has the energy 
\begin{equation}
\label{eight}E_0^{(0)}=-\frac{Z^2}2-\frac{Z_1^2}2=-Z^2+Z-1/2\;\;\;. 
\end{equation}
One can check easily that this energy is very close to the actual
ground-state energy of the two-electron atoms (cations).\ For example, we
have $E_0^{(0)}=-68\;eV$ as compared with the experimental value $E_0^{\exp
}=-79\;eV$ for the $He$ atom ($Z=2$), and $E_0^{(0)}=-176.8\;eV$, as
compared with $E_0^{\exp }=-198\;eV$ for the $Li^{+}$ cation. The remaining
interaction $V$ can be written as 
\begin{equation}
\label{nine}V=\sum_{l=1}^\infty \frac{r^l}{R^{l+1}}P_l\left( \cos \Omega
\right) =\sum_{l=1}^\infty \sum_{m=-l}^l\frac{4\pi }{2l+1}\cdot \frac{r^l}{%
R^{l+1}}\cdot Y_{lm}\left( \theta ,\varphi \right) Y_{lm}^{*}\left( \Theta
,\Phi \right) \;\;\;, 
\end{equation}
where $P_l$ are the Legendre polynomials, $Y_{lm}$ are the spherical
harmonics, $\Omega $ is the angle between ${\bf r}$ and ${\bf R}$ , and $%
{\bf r=(r,}\theta ,\varphi ),\;{\bf R=(R,}\Theta ,\Phi )$. This interaction
can, in principle, be treated perturbationally. However, the main
contribution to this interaction comes from those space regions where $r\sim
R$, so that the perturbation series converges extremely slowly.\
Consequently, we shall account for its main effect in another way.

Analyzing its expression given by $\left( 7\right) $ we see easily that $V$
has a minimum value at $r=R$ and $\Omega =\pi $; this value is given by $%
V=-1/2r$. We shall assume, therefore, that the ground-state wavefunction
given above is slightly distorted in such a way as to allow the second
electron to take full advantage of this minimum of energy. This amounts to
correcting $E_0^{(0)}$ by $\overline{V}=-1/2\overline{r}_0$, where $%
\overline{r}_0$ is the average value of the electron distances to the
nucleus $\overline{r}_1=\overline{r}_2=\overline{r}_0$ over the
hydrogen-like configuration $(100)\times (100)$ given above.\ We give here
the expression of this average radius%
$$
\overline{r}_0=C_0^2\cdot 16(ZZ_1)^3\int dr_2\cdot r_2^2\cdot
(\int_0^{r_2}dr_1\cdot r_1^3\cdot e^{-2Zr_1}\cdot e^{-2Z_1r_2}+ 
$$
\begin{equation}
\label{ten}+\int_{r_2}^\infty dr_1\cdot r_1^3\cdot e^{-2Z_1r_1}\cdot
e^{-2Zr_2})\;\;\;. 
\end{equation}
The ground-state energy is therefore 
\begin{equation}
\label{eleven}E_0=E_0^{(0)}-1/2\overline{r}_0\;\;\;, 
\end{equation}
and we remark that $-1/2\overline{r}_0$ is a classical correction,
corresponding to the principle of minimization of the energy for the
ground-state. In particular, we note that the first-order correction of $V$,
as given by $\left( 9\right) $, within the quantum-mechanical perturbation
theory vanishes for the ground-state.\ Within the present picture the
ground-state of the two-electron atom looks as being made of two electrons,
one moving inside, and the other outside an imaginary sphere, and changing
continuously their places through ${\bf r}_1\cong -{\bf r}_2$. Remark that
further corrections like zero-point oscillations around ${\bf r}_2=-{\bf r}%
_1 $ done for the $H^{-}$ anion would be inappropriate here since the second
electron moves everywhere (independently) outside the $r_2=r_1$ sphere in
the Coulomb field of charge $Z_1$. The normalization constants $C_0$, the
average radii $\overline{r}_0$ and the ground-state energies $E_0$ given by $%
\left( 11\right) $ are computed in {\it Table 1} for $Z=2$ through $11$, and
compared with the experimental values of the ground-state energies $%
E_0^{\exp }$. One can se that the agreement is very good, the errors being
less than $0.6\%$.

Within the present approximation the full wavefunctions of the ground-state
of the two-electron atom, including the spin degrees of freedom, may be
written (up to a normalization constant) as 
\begin{equation}
\label{twelve}\Psi _0({\bf r}_1,{\bf r}_2)=\left[ \alpha (1)\beta (2)-\alpha
(2)\beta (1)\right] \times \left\{ 
\begin{array}{c}
\Psi _{100}^{(Z)}( 
{\bf r}_1)\Psi _{100}^{(Z_1)}({\bf r}_2)\;\;,\;r_1<r_2\;, \\ \Psi
_{100}^{(Z)}({\bf r}_2)\Psi _{100}^{(Z_1)}({\bf r}_1)\;\;,\;r_1>r_2\;, 
\end{array}
\right\} \;, 
\end{equation}
for the spin-singlet state, and 
\begin{equation}
\label{thirteen}\Psi _0({\bf r}_1,{\bf r}_2)=\left\{ 
\begin{array}{c}
\alpha (1)\alpha (2) \\ 
\alpha (1)\beta (2)+\alpha (2)\beta (1) \\ 
\beta (1)\beta (2) 
\end{array}
\right\} \times \left\{ 
\begin{array}{c}
\Psi _{100}^{(Z)}( 
{\bf r}_1)\Psi _{100}^{(Z_1)}({\bf r}_2)\;\;,\;r_1<r_2\;, \\ -\Psi
_{100}^{(Z)}({\bf r}_2)\Psi _{100}^{(Z_1)}({\bf r}_1)\;\;,\;r_1>r_2\;, 
\end{array}
\right\} \;, 
\end{equation}
for the spin-triplet state, where $\alpha ,\;\beta $ are the spin-up and
spin-down wavefunctions.\ The wavefunctions written above are antisymmetric,
such as to satisfy the Pauli exclusion principle. Within the present
approximation these wavefunctions correspond to the same energy, so that the
para- and the ortho-atoms are degenerate.\ However, a closer inspection of $%
\left( 12\right) $, for example, tells us that the orbital of this
wavefunction, though continuous at ${\bf r}_1={\bf r}_2$ has discontinuous
derivatives at these points; and, similarly, the orbital wavefunction given
by $\left( 13\right) $ is discontinuous at ${\bf r}_1={\bf r}_2$. The exact
orbitals will be continuous, and will have continuous derivatives at ${\bf r}%
_1={\bf r}_2$, which implies that the spin-triplet orbital will have a
sudden variation on passing through zero at ${\bf r}_1={\bf r}_2$. This
indicates that the ortho-atom will have a higher energy, so that we may
conclude that the ground-state corresponds to a spin-singlet (para-atom).

We remark also that the wavefunctions given by $\left( 12\right) $ and $%
\left( 13\right) $ are,essentially, correlated wavefunctions, or genuine
two-electron wavefunctions.\ Indeed, one may define, for example, 
\begin{equation}
\label{fourteen}\varphi _2(1)=\left\{ 
\begin{array}{c}
\Psi _{100}^{(Z)}( 
{\bf r}_1)\;\;\;,\;r_1<r_2\;, \\ \Psi _{100}^{(Z_1)}({\bf r}%
_1)\;\;\;,\;r_1>r_2\;, 
\end{array}
\right\} \;\;, 
\end{equation}
and write the spin-singlet orbital of the ground-state as $\varphi
_2(1)\varphi _1(2)$; in spite of certain appearances, this is a correlated,
two-electron wavefunction, since, for example, according to its definition $%
\left( 14\right) $, the waefunction $\varphi _2(1)$, as function of ${\bf r}%
_1$, depends on ${\bf r}_2$.

We pass now to the lowest-excited states of the two-electron atoms
(cations). with nuclear charge $Z\geq 2$. It is easily to check that, to the
zeroth order approximation, the lowest-excited states correspond to the
hydrogen-like configurations $(100)\times (200)$ and $(100)\times (21m)$, 
{\it i.e.} the wavefunctions are $\Psi _{100}^{(Z)}({\bf r})\Psi
_{200}^{(Z_1)}({\bf R})$ and $\Psi _{100}^{(Z)}({\bf r})\Psi _{21m}^{(Z_1)}(%
{\bf R})$, according to the notations introduced above. The corresponding
energy is 
\begin{equation}
\label{fifteen}E_1=-\frac{Z^2}2-\frac{Z_1^2}8=-\frac{5Z^2}8+\frac Z4-\frac
18\;\;\;. 
\end{equation}
Indeed, one can check easily that the other alternative, corresponding to
the configurations $(200)\times (100),\;(21m)\times (100)$, has a much
higher energy.\ One can also check straightforwardly that the energies given
by $\left( 15\right) $ are already very close to the experimental values, as
one can see in {\it Table 2}, for $Z=2\;(He)$ and $Z=3\;(Li^{+})$; the
experimental values for larger $Z$, where available, have no longer been
included in {\it Table 2} as they correspond clearly to much higher-excited
states; for example, $E^{\exp }=-119.6\;eV$ for the $Be^{2+}$ cation,\cite
{Natl} while $\left( 15\right) $ gives $E=-248.2\;eV$ for $Z=4$. In {\it %
Table 2} there have also been included the normalization constants $C_1$ for
the configuration $(100)\times (21m)$, as well as the average radii $%
\overline{r}_1$ over this state for $Z=2$ through $11$. We shall remark,
first, that a correction of the type $-1/2\overline{r}_1$, as in the case of
the ground-state energy, is not appropriate here, since the excited states
are only stationary states, but they do not, of course, minimize the energy,
as the ground-state does. One may check that including $-1/2\overline{r}_1$
in the energy computed above would, indeed, result in worse numbers, i.e.
there would be a larger discrepancy with respect to the experimental values.
Secondly, we remark that although the two configurations $(100)\times (200)$
and $(100)\times (21m)$ are degenerate in this approximation, the former is
not orthogonal to the ground-state, as the latter is.\ An orthogonalization
procedure will push this state toward higher energies, so that the
lowest-excited state corresponds very closely to the $(100)\times (21m)$
configuration.\ Thirdly, we may also remark that an improved approximation
to the lowest excited level may proceed by standard perturbation
calculations applied to the interaction $V$ given by $\left( 9\right) $,
keeping in mind that we have to include those states that conserve the total
angular momentum.\ In the case of $(100)\times (21m)$ this is $L=1$, and the
interaction $V$ couples this state to $(21m)\times (100),\;(21m)\times
(200),\;(32m^{^{\prime }})\times (21m)$, etc. As we have said, the
perturbation series converges, however, very slowly, as the main
contributions come from $r\sim R$. And finally, let us remark that the
continuity of the two-particle orbitals at ${\bf r}_1={\bf r}_2$, and of
their derivatives, brings an additional splitting in energy between the
spin-singlet and spin-triplet states, as discussed above.

Obviously, the above scheme of electronic configurations may be extended to
multi-electron atoms.\ For example, for an atom with $N$ electrons, we may
set up (to the zeroth order approximation) the configuration $(nlm)\times
(n^{^{\prime }}l^{^{\prime }}m^{^{\prime }})\times (n^{^{\prime \prime
}}l^{^{\prime \prime }}m^{^{\prime \prime }})\times ...$ etc, where each
paranthesis denotes a hydrogen-like orbital corresponding, respectively, to
a nuclear charge $Z,\;Z_1=Z-1,\;Z_2=Z-2$, etc, up to $Z_{N-1}=Z-(N-1)$. The
remaining interaction of the type $V$ given in $\left( 7\right) $ can then
be minimized classically for the ground-state, according to the procedure
described above.\ Similar electronic configurations may also be used as
starting points for the excited states.\ Obviously, we have, in this
picture, correlated, multi-electronic wavefunctions, though each electron
moves in a central field, corresponding, however, to different nuclear
charges. One may infere that, within this picture, the atoms are more
''rarefied'', leading to increased transition probabilities, as compared
with those corresponding to all electrons moving in a unique nuclear charge.
Of course, the first thing to be done for checking the validity of such a
scheme of electronic configurations would be that of trying to account for
the periodicity of the chemical elements, i.e. the counterpart of the
closed-shell assumption of the self-consistent field model of the atom.

\newpage\ 

{\it Table 1}%
$$
\begin{array}{ccccc}
Z & C_0 & \overline{r}_0 & -E_0(eV) & -E_0^{\exp }(eV) \\ 
2\;(He) & 0.795 & 1.177 & 79.554 & 78.98 \\ 
3\;(Li^{1+}) & 0.856 & 0.650 & 197.713 & 198 \\ 
4\;(Be^{2+}) & 0.889 & 0.452 & 370.087 & 371.5 \\ 
5\;(B^{3+}) & 0.910 & 0.347 & 596.809 & 599.3 \\ 
6\;(C^{4+}) & 0.925 & 0.282 & 877.91 & 881.6 \\ 
7\;(N^{5+}) & 0.935 & 0.237 & 1.213\cdot 10^3 & 1.218\cdot 10^3 \\ 
8\;(O^{6+}) & 0.943 & 0.205 & 1.603\cdot 10^3 & 1.609\cdot 10^3 \\ 
9\;(F^{7+}) & 0.949 & 0.180 & 2.048\cdot 10^3 & 2.055\cdot 10^3 \\ 
10\;(Ne^{8+}) & 0.954 & 0.161 & 2.546\cdot 10^3 & 2.556\cdot 10^3 \\ 
11\;(Na^{9+}) & 0.958 & 0.145 & 3.099\cdot 10^3 & 3.11\cdot 10^3 
\end{array}
$$

{\it Table 2}%
$$
\begin{array}{ccccc}
Z & C_1 & \overline{r}_1 & -E_1(eV) & -E_1^{\exp }(eV) \\ 
2\;(He) & 0.709 & 2.907 & 57.8 & 59.16 \\ 
3\;(Li^{1+}) & 0.712 & 1.558 & 136 & 138.98 \\ 
4\;(Be^{2+}) & 0.714 & 1.092 & 248.2 & - \\ 
5\;(B^{3+}) & 0.715 & 0.853 & 394.4 & - \\ 
6\;(C^{4+}) & 0.717 & 0.708 & 574.6 & - \\ 
7\;(N^{5+}) & 0.718 & 0.610 & 788.8 & - \\ 
8\;(O^{6+}) & 0.718 & 0.539 & 1.037\cdot 10^3 & - \\ 
9\;(F^{7+}) & 0.719 & 0.485 & 1.319\cdot 10^3 & - \\ 
10\;(Ne^{8+}) & 0.719 & 0.444 & 1.635\cdot 10^3 & - \\ 
11\;(Na^{9+}) & 0.720 & 0.410 & 1.986\cdot 10^3 & - 
\end{array}
$$

\newpage\ 

Table captions

{\it Table 1}

The normalization constants $C_0$, the average radii $\overline{r}_0$, the
ground-state energies $E_0$ as given by $\left( 11\right) $ and the
experimental values $E_0^{\exp }$ of the ground-state energies\cite{Natl}
for the ground-state configuration $(100)\times (100)$ of the two-electron
atoms (cations), as functions of the nuclear charge $Z$.

{\it Table 2}

The normalization constants $C_1$, the average radii $\overline{r}_1$, the
energies $E_1$ as given by $\left( 15\right) $ for the lowest-excited states
corresponding to the configuration $(100)\times (21m)$, as well as the
experimental values $E_1^{\exp }$ for the lowest levels\cite{Natl} of the
two-electron atoms (cations), as functions of the nuclear charge $Z$; the
experimental values for $Z>3$ have not been included as they correspond to
higher excited levels.

\end{document}